\documentclass{aa}
\begin{document}
\newcommand\bsec{\hbox{$.\!\!{\arcsec}$}}
\newcommand\rsec{\hbox{$.\!\!{\rm ^s}$}}
\newcommand\RA[4]{#1$^{\rm h}$#2$^{\rm m}$#3\rsec#4}
\newcommand\DEC[3]{#1$^{\circ}$#2\arcmin#3\arcsec}
\newcommand\Msolar{$M_\odot$} 
\newcommand\teff{$T_{\rm eff}$} 
\newcommand\logt{$ \log\,T_{\rm eff}$}
\newcommand\logg{$ \log\,g$}
\newcommand\loghe{$ \log{\frac{n_{\rm He}}{n_{\rm H}}}$}
\thesaurus{05 (08.05.1; 08.06.3; 08.08.2; 10.07.3 NGC~6388; 10.07.3 NGC~6441)}
\title{Blue horizontal branch stars in metal-rich globular clusters.
 I. NGC~6388 and NGC~6441\thanks{Based on observations collected at the 
 European Southern Observatory (ESO~N$^{\b{o}}$~61.E-0361)} }
\author{S.~Moehler\inst{1}
 \and A.~V.~Sweigart\inst{2} \and 
 M.~Catelan\inst{2}\fnmsep\thanks{Hubble Fellow.}\fnmsep\thanks{Visiting
 Scientist, Universities Space Research Association.}}
\offprints{S.~Moehler}
\institute {Dr. Remeis-Sternwarte, Astronomisches Institut der Universit\"at
 Erlangen-N\"urnberg, Sternwartstr. 7, 96049 Bamberg, Germany\\
 e-mail: ai13@sternwarte.uni-erlangen.de
 \and NASA\,Goddard Space Flight Center, Code 681, Greenbelt, 
 MD 20771, USA \\
 e-mail: sweigart@bach.gsfc.nasa.gov, catelan@virginia.edu
}
\date{submitted April 9, 1999; revised version July 29, 1999}
\titlerunning{Blue HB stars in NGC~6388 and NGC~6441}
\maketitle

%%%%%%%%%%%%%%%%%%%%%%%%%%%%%%%%%%%%%%%%%%%%%%%%%%%%%%%%%%%%%%%%%%%%%%%%%
%% FOLLOWING TYPESETTING RULES SET OUT IN "ASTRONOMY AND ASTROPHYSICS %%
%% INSTRUCTIONS FOR AUTHORS" -- ASTRON. ASTROPHYS. 341 (1) 1999 %% 
%%%%%%%%%%%%%%%%%%%%%%%%%%%%%%%%%%%%%%%%%%%%%%%%%%%%%%%%%%%%%%%%%%%%%%%%%

\begin{abstract}
We report the first results of an ongoing
spectroscopic survey of blue horizontal branch stars in the metal-rich (${\rm
[Fe/H]} \simeq -0.5$) globular clusters NGC~6388 and NGC~6441. Based on data
obtained with the ESO--{\em New Technology Telescope} (NTT), we provide
gravities and temperatures for four stars in NGC~6388 and three stars in
NGC~6441. These results are marginally inconsistent
with the predictions of canonical evolutionary theory, but disagree 
strongly with all non-canonical scenarios that explain the sloped 
horizontal branches seen in the colour-magnitude diagrams of these clusters.
\end{abstract}


\section{Introduction\label{bulge_sec_intro}} 

The metal-rich globular clusters play a fundamental r\^ole in determining the
formation history of our Galaxy (e.g. Minniti \cite{minn95}, \cite{minn96}; 
Ortolani et al. \cite{orto95}; Zinn \cite{zinn96}; Barbuy et al. \cite{babi98}; 
Rich \cite{rich98}). Moreover, they provide a crucial template for interpreting
the integrated spectra of elliptical galaxies and for understanding the
evolution of old metal-rich stellar systems. The horizontal branch (HB) stars
in the metal-rich clusters are particularly important, since they provide a
standard candle for determining distances (and hence ages) and are
believed to be the major contributors to the UV-upturn phenomenon 
(e.g. Caloi \cite{calo89}; Greggio \& Renzini 1990, 1999; Bressan et al.
\cite{brch94}; Dorman et al. \cite{dooc95}; Yi et al. \cite{yide98}).

Recent {\em Hubble Space Telescope} (HST) observations have found that the 
metal-rich globular clusters NGC~6388 (C1732--447) and NGC~6441 (C1746--370) 
(${\rm [Fe/H]} = -0.60$ and $-0.53$, respectively; Harris \cite{harr96}) 
contain an unexpected population of hot HB stars and therefore exhibit the
well-known second parameter effect (Rich et al. \cite{rich97}). Most
surprisingly, the mean HB luminosity at the top of the blue HB tail is roughly
0.5~mag brighter in $V$ than the red HB ``clump," which itself is strongly
sloped as well. Differential reddening cannot be the cause of these sloped HB's
(Piotto et al. \cite{piot97}; Sweigart \& Catelan \cite{swca98}; Layden et al.
\cite{lari99}). 

The second parameter effect has often been attributed to differences in age or
mass loss on the red giant branch (RGB). However, canonical HB simulations 
show that increasing the assumed age or RGB mass loss moves an HB star 
blueward in the $V$,~$B-V$ plane but does not increase its luminosity. Thus 
some other second parameter(s) must be causing the sloped HB's in NGC~6388 and
NGC~6441 (Sweigart \& Catelan \cite{swca98}, hereafter SC98). 

Three non-canonical scenarios have been suggested to explain the sloped HB's 
and long blue HB tails in these metal-rich globular clusters (SC98):
\begin{enumerate}
\item {\bf High cluster helium abundance scenario}:
Red HB stars evolve along blue loops during most of their HB lifetime. For
larger than ``standard" helium abundances $Y$, these loops
become considerably longer, reaching higher effective temperatures and
deviating more in luminosity from the zero-age HB (ZAHB). 
If the cluster stars form with sufficiently high
$Y$ ($\stackrel{>}{_\sim}0.4$), the HB will slope upward (Catelan \& de Freitas
Pacheco \cite{cadf96}), as observed in NGC~6388 and NGC~6441. However, this
scenario also predicts a much larger value for the number ratio $R$ of HB to RGB
stars than the value recently obtained by Layden et al. (\cite{lari99}) for
NGC~6441. Thus a high primordial helium abundance seems to be
ruled out as the cause of the sloped HB's and will not be 
considered further in this paper.

\item {\bf Rotation scenario}:
Rotation during the RGB phase can delay the helium flash, thereby increasing
both the final helium-core mass and the amount of mass loss near the tip of the
RGB. The net effect is to shift a star's HB location towards higher effective
temperatures and luminosities, depending on the amount of rotation. This
scenario predicts a sloped HB with the shift towards higher luminosities (and
hence lower gravities) increasing with effective temperature (Rood \& Crocker
\cite{rood89}).

\begin{table*}
\caption[]{Target list. Positions and photometry are from Piotto
(priv. comm.)\label{bulge_targ}}
\begin{tabular}{lr|ll|rr}
\hline
Cluster & Star & $\alpha_{2000}$ & $\delta_{2000}$ & $V$ & $B-V$ \\
        &      &  &  & [mag] & [mag] \\
\hline
NGC~6388 & WF2--42 & \RA{17}{36}{21}{4} & \DEC{$-$44}{42}{56} & 17.440 
& $+$0.334 \\
NGC~6388 & WF3--14 & \RA{17}{36}{23}{5} & \DEC{$-$44}{43}{11} & 16.914 
& $+$0.333 \\
NGC~6388 & WF3--15 & \RA{17}{36}{24}{1} & \DEC{$-$44}{43}{15} & 16.896 
& $+$0.432 \\
NGC~6388 & WF4--20 & \RA{17}{36}{20}{9} & \DEC{$-$44}{45}{10} & 17.183 
& $+$0.341 \\
\hline
NGC~6441 & WF2--24 & \RA{17}{50}{08}{5} & \DEC{$-$37}{04}{09} & 17.567 
& $+$0.386 \\
NGC~6441 & WF3--16 & \RA{17}{50}{06}{2} & \DEC{$-$37}{02}{50} & 17.683 
& $+$0.558 \\
NGC~6441 & WF3--17 & \RA{17}{50}{04}{5} & \DEC{$-$37}{02}{20} & 17.643 
& $+$0.499 \\
\hline
\end{tabular}
\end{table*}

\item {\bf Helium-mixing scenario}:
The observed abundance variations in globular cluster RGB stars show that these
stars are able to mix nuclearly processed material from the vicinity of the
hydrogen-burning shell out to the surface (e.g. Kraft \cite{kraf94}). In
particular, the observed Al enhancements indicate that the mixing can penetrate
into the H-shell, thus leading to the dredge-up of helium. The resulting
increase in the envelope helium abundance produces an HB morphology that slopes
upward towards brighter luminosities with increasing effective temperature
(Sweigart \cite{swei99}; SC98). 
The shift towards lower than canonical
gravities due to helium mixing reaches a maximum between 10,000~K and 20,000~K.
No shift is predicted for the hottest HB stars ($>$20,000~K), where the 
H-shell is inactive
and the luminosity is therefore unaffected by helium mixing (Sweigart
\cite{swei99}). This scenario might also help to explain the low gravities of
the blue HB stars found in several metal-poor globular clusters (Moehler
\cite{moeh99} and references therein), although Grundahl et al. (\cite{grca99})
have recently provided some caveats about it. They do note, however, that
``helium mixing stands out as the best candidate to explain the anomalous HB
morphology of the metal-rich globular clusters NGC~6388 and NGC~6441". 
\end{enumerate}

These scenarios make different predictions for the surface gravities of HB 
stars which can be tested observationally. We started a program to obtain 
spectroscopic observations with this goal in mind and report here on our first
results. In Sect.~\ref{bulge_sec_obs} we describe our observations and the 
employed data reduction techniques; in Sect.~\ref{bulge_sec_par}
the procedure adopted to derive the
atmospheric parameters of the programme stars is outlined. Finally, we discuss
our results and their consequences in Sect.~\ref{bulge_sec_disc}.

\section{Observations and Data Reduction\label{bulge_sec_obs}}

We selected our targets from the HST {\em Wide Field and Planetary Camera--2} 
(WFPC2) images of Rich et al. (\cite{rich97}) which were convolved with a 
``seeing'' of $1''$ in order to allow the selection of stars suitable for {\em
ground-based} spectroscopic observations. We took spectra of four stars in
NGC~6388 and three stars in NGC~6441 (see Table~\ref{bulge_targ}). Unfortunately, due to 
rather mediocre weather conditions we could not observe more stars during our 
{\em New Technology Telescope} (NTT) observing run. 

We used only the blue channel of the {\em ESO Multi-Mode Instrument} (EMMI) at
the NTT, since the use of the two-channel mode leads to destructive
interference around the H$_\beta$ line. We obtained medium resolution spectra
to measure Balmer line profiles and helium line equivalent widths. We used
grating \#4 (72~\AA$\,$mm$^{-1}$) and a slit width of 1\bsec0, keeping the slit
at parallactic angle for all observations. Seeing values varied between 0\bsec8
and 1\bsec5. The CCD was a Tek~$1024 \times 1024$ chip with (24$\,\mu$m)$^2$
pixels, a read-out-noise of 5.7~e$^-$ and a gain of 2.84 e$^-$ADU$^{-1}$ (where
as usual ``ADU" stands for analog-to-digital-units).

For calibration purposes we observed each night ten bias frames and ten dome
flat-fields with a mean exposure level of about 10,000~ADU each. In the
beginning of each night we took a long HeAr exposure for wavelength 
calibration. Due to the long exposure times for the Ar spectrum we only 
obtained He calibration spectra during the night (before and after each 
science observation) and determined the offsets relative to the HeAr spectrum
(from which the dispersion relation was derived) by correlating the two
spectra. For the second night (and part of the third night) we could not obtain
{\em any} wavelength calibration spectra due to technical problems. To correct 
any
spectral distortions we used HeAr spectra from the first night for those data.
The FWHM of the HeAr lines was measured to be 5.67$\pm$0.2~\AA\ and was 
used as instrumental resolution.
As flux standard stars we used LTT~7987 and EG~274.

We averaged the bias frames of the three nights and used their mean value 
instead of the whole frames as there was no spatial or temporal variation 
detectable. To correct the electronic offset we adjusted the mean bias by the
difference between the mean overscan value of the science frame and that of the
bias frame. The dark currents were determined from several long dark frames and
turned out to be negligible ($5.6 \pm 4.2$ $\frac{e^-}{hr px}$). 

The flat fields were averaged separately for each night, since we detected a
slight variation in the fringe patterns of the flat fields from one night to
the next (below 5\%). We averaged the flat fields and then determined the 
spectral energy distribution of the flat field lamp by averaging the flat 
fields along the spatial axis. This one-dimensional ``flat field spectrum" was
then heavily smoothed and used afterwards to normalize the dome flats along
the dispersion axis. 

For the wavelength calibration we fitted 3$^{\rm rd}$-order polynomials to the
dispersion relations of the HeAr spectra. They were adjusted for eventual 
offsets by cross-correlating them with the He spectra belonging to the science
frame. We rebinned the frames two-dimensionally to constant wavelength steps.
Before the sky fit the frames were smoothed along the spatial axis to erase
cosmics in the background. To determine the sky background we had to find
regions without any stellar spectra, which were sometimes not close to the
place of the object's spectrum. Nevertheless the flat field correction and
wavelength calibration turned out to be good enough that a linear fit to the
spatial distribution of the sky light allowed the sky background at the
object's position to be reproduced with sufficient accuracy. This means in our
case that after the fitted sky background was subtracted from the unsmoothed
frame we do not see any absorption lines caused by the predominantly red stars
of the clusters. The sky-subtracted spectra were extracted using Horne's
(\cite{horn86}) algorithm as implemented in MIDAS (Munich Image Data Analysis
System). 

Finally the spectra were corrected for atmospheric extinction using the data 
of T\"ug (\cite{tueg77}). The data for the flux standard stars were taken from
Hamuy et al. (\cite{hamu92}) and the response curves were fitted by splines. 

\begin{figure}
\vspace{11.6cm}
\includegraphics{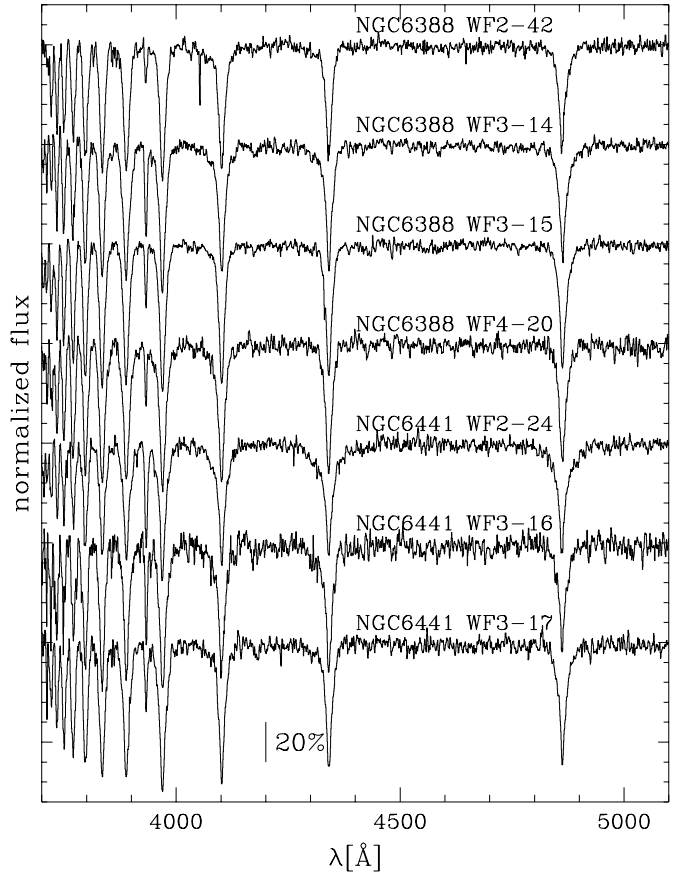}
\caption{Normalized spectra for the programme stars\label{bulge_spectra}}
\end{figure}

\section{Atmospheric Parameters\label{bulge_sec_par}}

To derive effective temperatures, surface gravities and helium abundances we
fitted the observed Balmer and helium lines with stellar model atmospheres.
Beforehand we corrected the observed spectra for wavelength shifts 
introduced by radial velocities and the lack of appropriate wavelength 
calibration frames (see Sect.~\ref{bulge_sec_obs}), derived from
the positions of the Balmer lines. The individual
spectra for each star were then co-added and normalized
by eye and are plotted in Fig.~\ref{bulge_spectra}. For the 
normalization we assumed that the observed noise is mostly caused by 
photon noise. However, to account for the 
possibility of faint unresolved metal lines we tried to place the 
continuum line at the upper region of the scatter.

To establish the best fit we used the routines developed by Bergeron et al.
(\cite{besa92}) and Saffer et al. (\cite{saff94}), which employ a $\chi^2$ 
test. In addition the fit program normalizes model spectra {\em and} observed 
spectra using the same points for the continuum definition.
The $\sigma$ necessary for the calculation of $\chi^2$ is estimated 
from the noise in the continuum regions of the spectra.

We computed model atmospheres using ATLAS9 (Kurucz \cite{kuru91}, priv. 
comm.) and used 
Michael Lemke's version\footnote{\scriptsize 
http://www.sternwarte.uni-erlangen.de/$^\sim$ai26/linfit/linfor.html} of the
LINFOR program (developed originally by Holweger, Steffen, and Steenbock at
Kiel University) to compute a grid of theoretical spectra which include the
Balmer lines H$_\alpha$ to H$_{22}$ and \ion{He}{I} lines. The grid covered the
range 7500~K~$\leq$~\teff~$\leq$~20,000~K, 2.5~$\leq$~\logg~$\leq$~5.0, 
$-3.0$~$\leq$~\loghe~$\leq$~$-1.0$, at metallicities of [M/H]~=~$-0.75$ and
[M/H]~=~$-0.50$. To check the effects of metallicity we fitted the spectra at
\loghe~=~$-1.0$ using model spectra with [M/H]~=~$-0.50$ and [M/H]~=~$-0.75$.
The fit results for the two metallicities were almost identical: \teff\ and
\logg, respectively, increased by at most 1\% and 0.03~dex from [M/H]~=~$-0.50$
to $-0.75$ (except for NGC~6441~WF3--16, where $\Delta\,$\teff\ and
$\Delta\,$\logg\ were 1.6\% and 0.07~dex, respectively). All further analysis
was therefore performed for a fixed metallicity of [M/H]~=~$-0.50$.

As none of the programme stars showed any He absorption lines we first fitted 
all spectra at two fixed helium abundances, \loghe~=~$-1.0$ and $-2.0$, using
the Balmer lines H$_\beta$ to H$_{10}$ (except H$_\epsilon$ to avoid 
contamination by the \ion{Ca}{II}~H line). When defining the fit 
regions we took care to extend them as far as possible while
avoiding any predicted strong \ion{He}{I} lines ($\lambda\lambda$ 4026~\AA,
4388~\AA, 4921~\AA) which might be present at the noise level although 
undetected by eye. Such lines would distort the fit: Fitting regions of 
strong \ion{He}{I} lines while keeping the helium abundance fixed
will result in erroneous results for \teff\ and 
\logg\ if the helium abundance of the model atmospheres is not identical 
with that of the star. The results are listed in 
Table~\ref{bulge_par}.

\begin{table*}
\caption{Atmospheric parameters and equivalent widths of the
\ion{Ca}{II} K line for the programme stars for an assumed [M/H]~=~$-0.5$.
For stars hotter than 9,500 K the first row gives the atmospheric 
parameters obtained by fitting the Balmer lines and the spectral region of the 
strongest \ion{He}{I} lines. For all stars
the first two of the four
rows at fixed helium abundance give the solutions on the hot side 
of the Balmer maximum for the cited helium abundances, while the last two 
rows give the solutions on the cool side of the Balmer maximum that were 
rejected because of the strength of the \ion{Ca}{II} K line. The errors are 
1$\sigma$ errors adjusted for $\chi^2\, > 1$ (see also 
Sect.~\ref{bulge_sec_par}).
A colon marks extrapolated values\label{bulge_par}}
\begin{tabular}{l|lrrrrrr|rr}
\hline
Star & $\chi^2$ & \teff & $\delta\,$\teff & \logg & $\delta\,$\logg & 
\loghe & $\delta\,$\loghe &
\multicolumn{2}{c}{$W_\lambda$~(\ion{Ca}{II}~K)} \\
 & & & & & & & & obs. & model\\
 & & [K] & [K] & [cm$\,$s$^{-2}$] & [cm$\,$s$^{-2}$] & & & [\AA] & [\AA] 
\\
\hline
 NGC~6388 WF2--42 
 & 4.34 & 12130 & 370 & 4.14 & 0.12 & $\le-$2.0 & 0.6 & 
 1.2 & 0.1\\
 & 4.78 & 12090 & 310 & 4.03 & 0.09 & $-$1.00 & & & \\
 & 4.78 & 12140 & 320 & 4.15 & 0.09 & $-$2.00 & & & \\
 & 4.75 &  7010 &  70 & 1.9: & 0.26 & $-$1.00 & & & 6.5\\
 & 4.73 &  7060 &  50 & 2.1: & 0.15 & $-$2.00 & & & \\
& & & & & & & & \\
 NGC~6388 WF3--14 
 & 4.12 & 12250 & 310 & 4.56 & 0.10 & $\le-$1.9 & 0.4 &
 3.1 & 0.1 \\
 & 3.86 & 12180 & 250 & 4.44 & 0.07 & $-$1.00 & & & \\
 & 3.85 & 12210 & 240 & 4.56 & 0.07 & $-$2.00 & & & \\
 & 3.82 &  7450 &  40 & 2.79 & 0.08 & $-$1.00 & & & 4.2\\
 & 3.83 &  7450 &  40 & 2.87 & 0.07 & $-$2.00 & & & \\
& & & & & & & & \\
 NGC~6388 WF3--15 
 & 2.94 &  9080 & 590 & 3.13 & 0.34 & $-$1.00 & & 1.7 & 0.7 \\
 & 2.94 &  9130 & 450 & 3.26 & 0.26 & $-$2.00 & & & \\
 & 2.94 &  9100 & 570 & 3.14 & 0.33 & $-$1.00 & & & \\
 & 2.94 &  9160 & 420 & 3.29 & 0.24 & $-$2.00 & & & \\
& & & & & & & & \\
 NGC~6388 WF4--20 
 & 2.18 &  9960 & 320 & 3.71 & 0.20 & $\le-$1.5 & 0.9 &
  1.4 & 0.3 \\
 & 1.82 &  9900 & 250 & 3.62 & 0.12 & $-$1.00 & & & \\
 & 1.83 &  9910 & 250 & 3.73 & 0.12 & $-$2.00 & & & \\
 & 1.79 &  8150 & 140 & 2.69 & 0.07 & $-$1.00 & & & 2.5 \\
 & 1.79 &  8130 & 130 & 2.77 & 0.06 & $-$2.00 & & & \\
& & & & & & & & \\
 NGC~6441 WF2--24 
 & 2.96 & 14360 & 420 & 5.2: & 0.11 & $\le-$1.8 & 0.4 &
 3.8 & 0.1 \\ 
 & 3.20 & 14460 & 320 & 5.1: & 0.08 & $-$1.00 & & & \\
 & 3.28 & 14400 & 350 & 5.2: & 0.08 & $-$2.00 & & & \\
 & 1.99 &  7570 &  20 & 5.0: & 0.11 & $-$1.00 & & & 4.1 \\
 & 2.01 &  7570 &  20 & 5.0: & 0.11 & $-$2.00 & & & \\
& & & & & & & & \\
 NGC~6441 WF3--16 
 & 2.61 & 12710 & 480 & 4.65 & 0.14 & $\le-$1.8 & 0.4 
 & 3.8 & 0.1 \\
 & 2.23 & 12630 & 360 & 4.53 & 0.10 & $-$1.00 & & & \\
 & 2.22 & 12630 & 360 & 4.63 & 0.10 & $-$2.00 & & & \\
 & 1.97 &  7360 &  40 & 3.14 & 0.15 & $-$1.00 & & & 4.2\\
 & 1.97 &  7370 &  40 & 3.23 & 0.15 & $-$2.00 & & & \\
& & & & & & & & \\
 NGC~6441 WF3--17 
 & 2.55 & 10910 & 310 & 4.04 & 0.15 & $\le-$1.0 & 0.5 &
 2.3 & 0.2 \\
 & 2.44 & 10920 & 250 & 4.05 & 0.10 & $-$1.00 & & & \\
 & 2.44 & 10930 & 250 & 4.16 & 0.10 & $-$2.00 & & & \\
 & 2.38 &  7720 &  70 & 2.61 & 0.05 & $-$1.00 & & & 3.7\\
 & 2.38 &  7720 &  70 & 2.71 & 0.05 & $-$2.00 & & & \\
\hline
\end{tabular}
\end{table*}

As the stars are rather cool it is not obvious from the spectra on which side 
of the Balmer maximum they lie. However, the strength of the \ion{Ca}{II}~K 
line can be used to distinguish between the ``hot'' and the ``cool'' solution:
the observed equivalent width of this line has stellar and interstellar 
contributions. As the clusters are highly reddened~--~$E(B-V) = 0.40$ 
and $0.44$
for NGC~6388 and NGC~6441, respectively (Harris \cite{harr96})~--~{\em the 
observed equivalent width has to be significantly larger than the one 
predicted by the model atmosphere} for [M/H]~=~$-0.5$. In 
Table~\ref{bulge_par} we list the
atmospheric parameters for all stars together with the measured and predicted
equivalent widths for the \ion{Ca}{II}~K line. {\em The comparison between
observed and predicted values places all ambiguous stars on the hot side of the
Balmer maximum.} 

\begin{figure*}
\vspace{13.cm}
\includegraphics{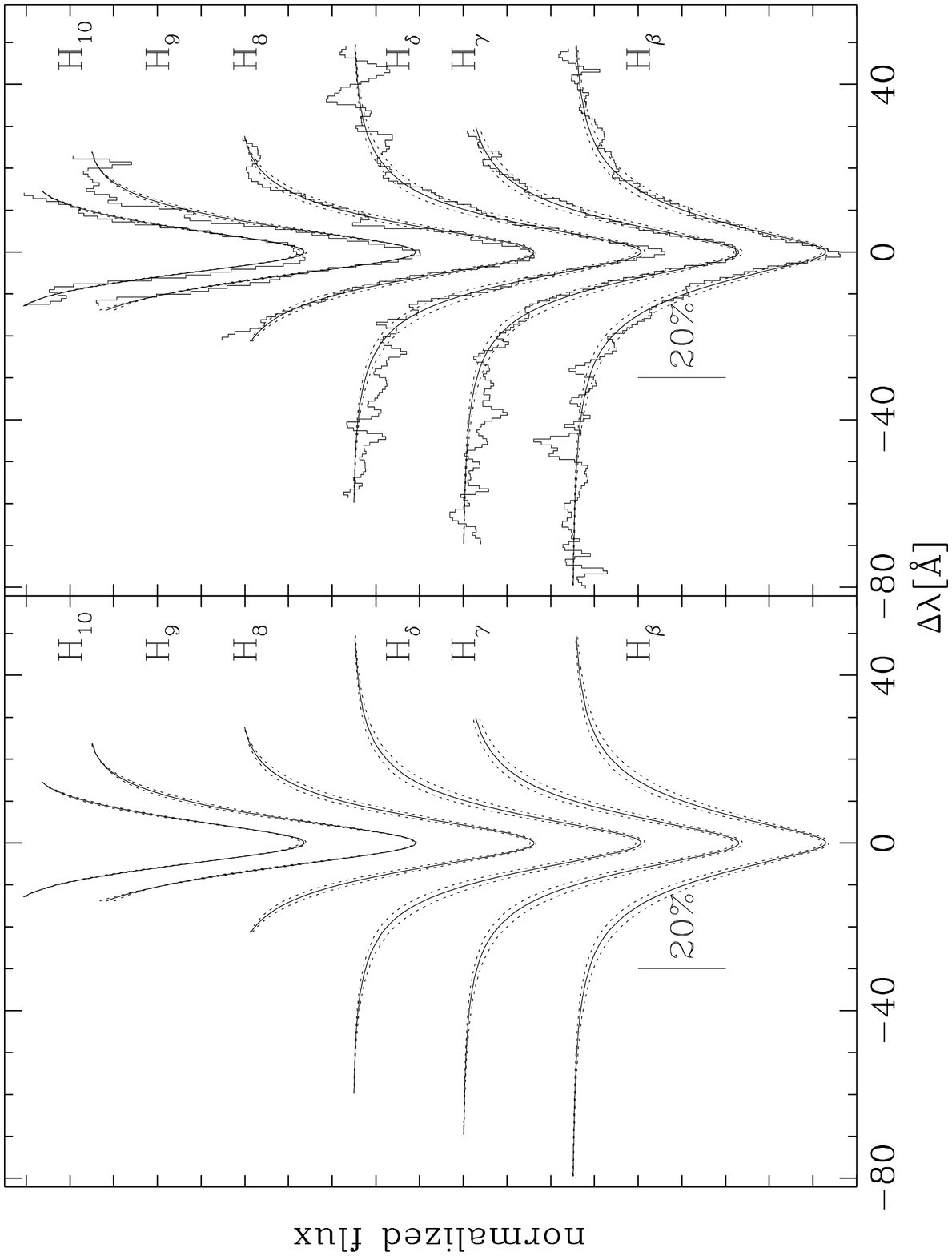}
\caption{Plots of model atmosphere spectra for \teff\ = 10,910~K and \logg\ 
= 4.04 (solid line), 3.84 (upper dashed line), 4.24 (lower dashed line). 
The model spectra were convolved by a Gaussian of 5.67~\AA\ FWHM to take 
into account the instrumental resolution of the observed data. Given are those 
ranges of the Balmer lines that were fitted to obtain effective temperatures 
and surface gravities. The asymmetric fit regions are due to the avoidance of 
\ion{He}{I} lines that were fitted separately (see Sect.~\ref{bulge_sec_par}). 
The right panel shows the observed line profiles of NGC~6441 WF3--17 in 
comparison\label{bulge_profile}}
\end{figure*}

For those stars that are hotter than 9,500~K we finally fitted the Balmer 
lines mentioned above together with those regions of the spectra where the
\ion{He}{I} lines $\lambda\lambda$~4026~\AA , 4388~\AA , 4471~\AA , and
4921~\AA\ are expected. This way we want to put at least an upper limit to the
He abundance even though no He lines are visible. As can be seen from 
Table~\ref{bulge_par}
this upper limit tends to be subsolar for the hotter stars, in agreement with
the results for blue HB stars in metal-poor clusters (e.g. Moehler et al.
\cite{moeh97}, \cite{moeh99a}; Behr et al. \cite{beco99}). 

\subsection{Possible errors and discrepancies}

\begin{enumerate}

\item {\bf Error estimates:\ }
The fit program gives 1$\sigma$ errors 
derived from $\Delta \chi^2$ = 2.71 (\teff , \logg) resp. 3.53 (\teff , 
\logg , \loghe). However, since $\chi^2$ is not close to 1 these errors 
will most likely underestimate the true errors. As the errors causing the 
large $\chi^2$ values are most probably systematic (see below) they are 
hard to quantify. We decided to get an estimate of their size by assuming 
that the large $\chi^2$ is solely caused by noise. Increasing $\sigma$ 
until $\chi^2$ = 1 then yields new formal errors for $\Delta \chi^2$ = 2.71
and 3.53, respectively, which are given in Table~\ref{bulge_par}.
We therefore adopt a general overall error in \logg\ of $\pm$0.2~dex and in 
\teff\ of 5\%. Fig.~\ref{bulge_profile} shows theoretical line profiles for the 
parameters derived for NGC~6441 WF3--17 and model atmosphere spectra for 
$\Delta$\logg = $\pm$0.2 dex. 

\item {\bf Large $\chi^2$:\ }
As can be seen from Table~\ref{bulge_par} the reduced $\chi^2$ is 
far from 1. We believe that these bad $\chi^2$ values are due to the low 
resolution of the data (5.7~\AA): At \teff\ = 11,000~K and \logg\ = 4 the 
FWHM of the theoretical Balmer lines are around 12~\AA . Thus the 
instrumental profile makes up a considerable part of the observed line 
profile. Therefore any deviations of the instrumental profile from a Gaussian 
(which is used to convolve the model atmospheres) will lead to a bad fit. 
Fortunately we could compare the effects for one blue HB star in NGC~6752, for 
which we have an NTT spectrum with the same setup as is used here and a 
spectrum from the ESO 1.52m telescope with a resolution of 2.6~\AA\
(for details see Moehler et al. \cite{moeh99b}). 
Performing a spectroscopic analysis as described above with model 
atmospheres of [M/H] = $-$1.5 we find the parameters listed in 
Table~\ref{bulge_tab_compa}. While the 
$\chi^2$ values for both fits differ considerably the resulting effective 
temperatures and surface gravities are still rather similar.

\item {\bf Continuum placement:\ }
Using the spectrum of NGC~6441 WF3--17 we checked the effects of 
a possible misplacement of the continuum: We now normalized the spectrum 
again, using only the uppermost resp. lowermost continuum points.
This amounted to changes of $-$1\% (upper continuum) resp. $+$3\% (lower 
continuum) in the normalized continuum level. Temperature, gravity and
$\chi^2$ were 
only slightly affected by these changes: In both cases the temperature 
increased by about 50~K, \logg\ by 0.03~dex, and $\chi^2$ by about 0.04. 
The effect of a misplaced continuum is mostly erased by the renormalization of 
observed {\em and} model spectrum before fitting, using the same continuum 
points for both.

\end{enumerate}

The effective temperatures from Table~\ref{bulge_par} and the $B-V$ colours 
from Table~\ref{bulge_targ} do not correlate well, esp. for the stars in 
NGC~6441. However, as the reddening towards these globular clusters
is relatively large [$E(B-V) \approx 0.4$~mag] and {\em may} vary by a few
hundredths of a magnitude on small scales (especially in the case of NGC~6441;
see Sect.~3.1 in Piotto et al. \cite{piot97}; for a more general discussion 
see Heitsch \& Richtler \cite{heri99}), we do not put too much emphasis
on this (apparent) inconsistency between spectroscopically derived temperatures
and observed (WFPC2) colours.

\begin{figure}
\vspace{11.6cm}
\includegraphics{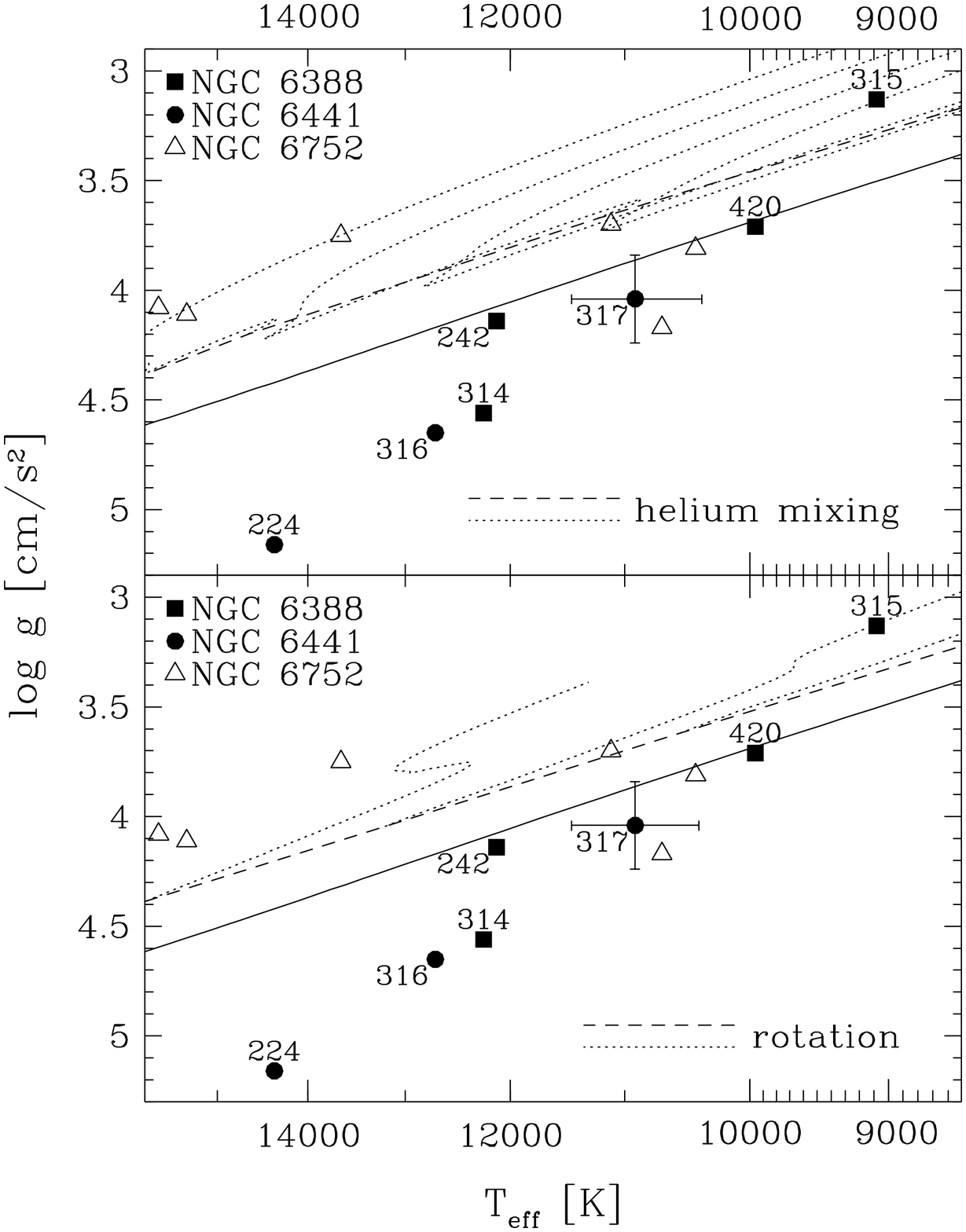}
\caption{
Comparison of the measured gravities and temperatures for our
programme stars in NGC 6388 and NGC 6441 against the HB models for
[M/H] = $-$0.5 from Sweigart \& Catelan (\cite{swca99}).  
The solid curve represents a canonical zero-age HB (ZAHB) for this metallicity. 
The ZAHB and evolutionary tracks for the sequences which reproduce
the observed HB slope in these globular clusters are indicated
by the dashed and dotted curves, respectively. 
The {\bf upper panel} shows the helium-mixed tracks, the {\bf
lower panel} shows the rotation tracks. The numbers of the stars 
refer to Table~\ref{bulge_par}. A representative error bar is plotted at
the location of the star NGC~6441 WF3--17.
In addition we show cool blue HB stars in NGC~6752, which were observed with 
the same instrumental setup during the same run
(Moehler et al. \cite{moeh99b})\label{bulge_tg}}
\end{figure}

\section{Discussion and Future Work\label{bulge_sec_disc}}

The atmospheric parameters derived for our programme stars are compared in
Fig.~\ref{bulge_tg} to both the canonical ZAHB 
and the non-canonical tracks for
[M/H]~=~$-0.5$ from Sweigart \& Catelan (\cite{swca98}, \cite{swca99}, 
helium-mixed resp. rotation, cf. Sect.~\ref{bulge_sec_intro}). 
Quite unexpectedly, the
studied stars tend to lie preferentially {\em below} the canonical
ZAHB\footnote{The difference between Fig.~\ref{bulge_tg} in this paper and 
Fig.~8 in
Moehler (\cite{moeh99}) results from using fitting ranges for the preliminary
analysis of the line profiles that did not extend far enough to include the
wings of the Balmer line profiles.}. 
This behaviour stands in marked contrast to the results for
blue HB stars in more metal-poor globular clusters, where lower than canonical
gravities are normally found (see Moehler \cite{moeh99} and references
therein).
To verify that the high gravities are not due to problems with the low 
resolution of the data, we also show results for cool blue HB stars in 
NGC~6752 that were observed during the same run with the same resolution
(Moehler et al. 
\cite{moeh99b}). It can be clearly seen that those stars follow the 
overall trend seen in NGC~6752 to show lower than expected surface 
gravities when fitted with metal-poor model atmospheres. If the high 
gravities found for the stars in NGC~6388 and NGC~6441 were caused by 
systematic errors in the data analysis the stars in NGC~6752 should exhibit 
the same effect.

\begin{table}
\caption[]{Results of spectroscopic analyses for star B~3253 in NGC~6752 
\label{bulge_tab_compa}}
\begin{tabular}{l|c|c}
\hline
 & NTT & ESO 1.52m\\
\hline
$\chi^2$ & 5.09 & 2.18 \\
\teff\ [K] & 13,630$\pm$260 & 13,810$\pm$140\\
\logg\ [cm s$^{-2}$] & 3.86$\pm$0.06 & 3.99$\pm$0.03 \\
\loghe & $-$2.2$\pm$0.19 & $-$2.7$\pm$0.14 \\
\hline
\end{tabular}
\end{table}

Considering the estimated errors in 
\teff\ and \logg\ the positions of NGC~6388 WF2--42, WF3--14, WF4--20, and 
NGC~6441 WF3--17, however, are 
consistent with the canonical ZAHB within 2$\sigma$. NGC~6388 WF3--15 could 
be a blue HB star already on its way to the asymptotic giant branch -- as 
we observed rather bright blue HB stars at the top of the blue tail
in both clusters we are 
biased towards stars evolving from the ZAHB towards higher luminosities.
NGC~6441 WF3--16 lies marginally more than 2$\sigma$ below the unmixed ZAHB,
but has also the lowest S/N of all observed stars. 
NGC~6441 WF2--24 lies 0.7~dex in \logg\ below the canonical 
ZAHB along the track of a 0.195~\Msolar\ helium white dwarf
(Driebe et al. \cite{drsc98}). In this case it would be a foreground object 
and most probably member of a binary system having undergone mass transfer, 
as the low mass single star precursors of helium white dwarfs evolve too 
slowly to reach this stage within a Hubble time. While helium white dwarfs 
evolve rather fast and the probability of finding one is therefore rather 
low, we could not find any other physical explanation for a star like
NGC~6441 WF2--24. As we are not able to derive radial velocities (see 
Sect.~\ref{bulge_sec_obs}) we cannot decide whether this star is a cluster 
member or not.

The derived gravities (except for NGC~6388 WF3--15) are {\em significantly} 
larger than those predicted by the non-canonical tracks 
that reproduce the upward sloping HB's in
the colour-magnitude diagrams of NGC~6388 and NGC~6441 (see 
Sect.~\ref{bulge_sec_intro}). 
The discrepancy would increase if the atmospheric 
abundances of the programme stars (and thereby their line profiles) were 
affected by the radiative levitation of metals (Grundahl et al. \cite{grca99}).
As shown by Moehler et al. (\cite{moeh99a}) accounting for this effect 
moves the parameters of blue HB stars in NGC~6752 to lower temperatures and/or 
higher gravities. However, the moderate resolution and S/N of the data 
discussed here do not allow abundances of the heavy elements to be 
estimated.

We do not have an explanation for this surprising result. All the scenarios
laid out by SC98 and discussed in 
Sect.~\ref{bulge_sec_intro}
predict anomalously low gravities for cluster blue HB stars within the
temperature range of our programme stars. As far as we are aware, there are no
alternative models capable of accounting for the sloped HB's seen in NGC~6388
and NGC~6441 without producing anomalously bright HB stars and hence low
gravities. In fact, recent analyses of RR Lyrae variables in NGC~6388 and
NGC~6441 (SC98; Layden et al. \cite{lari99}; 
Pritzl et al. \cite{prsm99}) {\em strongly} indicate that the HB's of these
globular clusters are substantially brighter than canonical models would
predict. Thus we face a conundrum: the non-canonical models which explain the
upward sloping HB's in these globular clusters are inconsistent with the
derived gravities.  

It is clear that further spectroscopic observations of larger 
(statistically more significant) samples of blue HB
stars in NGC~6388 and NGC~6441 are urgently needed to verify the present
results.

\acknowledgements

We want to thank the staff of the ESO La Silla observatory for their support 
during our observations and U. Heber, W.B.~Landsman, R. Napiwotzki, and 
S. Ortolani for their help in improving this paper. 
%M.C. would like to thank W.B. Landsman for useful discussions. 
Thanks go also to an anonymous referee for valuable remarks. 
S.M. acknowledges financial support from the DARA under grant 50~OR~96029-ZA.
Support for M.C.
was provided by NASA through Hubble Fellowship grant HF--01105.01--98A awarded
by the Space Telescope Science Institute, which is operated by the Association
of Universities for Research in Astronomy, Inc., for NASA under contract
NAS~5-26555.

%%%%%%%%%%%%%%%%%%%%%%%%%%%%%%%%%%%%%%%%%%%%%%%%%%%%%%%%%%%%%%%%%%%%%%%%%
%% FOLLOWING TYPESETTING RULES SET OUT IN "ASTRONOMY AND ASTROPHYSICS %%
%% INSTRUCTIONS FOR AUTHORS" -- ASTRON. ASTROPHYS. 341 (1) 1999; %%
%% JOURNAL ACRONYMS CAN BE FOUND AT THE FOLLOWING TWO WEB SITES: %%
%% http://adsdoc.harvard.edu/abs_doc/refereed.html %%
%% http://adsdoc.harvard.edu/abs_doc/non_refereed.html %%
%%%%%%%%%%%%%%%%%%%%%%%%%%%%%%%%%%%%%%%%%%%%%%%%%%%%%%%%%%%%%%%%%%%%%%%%%

\end{document}